\documentclass[amsmath,amssymb,aps,superscriptaddress,pra,twocolumn,longbibliography,10pt]{revtex4-2}

\usepackage[sort&compress]{natbib}
\usepackage{comment}
\usepackage{amsthm}
\usepackage{mathrsfs}
\usepackage{braket,soul}
\usepackage[utf8]{inputenc}
\usepackage{xcolor,graphicx} 
\usepackage{esint}
\newcommand\D{\!\operatorname{d}\!}
\providecommand{\x}{\mathbf{x}}

\providecommand{\q}{\mathbf{q}}
\providecommand{\p}{\mathbf{p}}
\providecommand{\y}{\mathbf{y}}
\providecommand{\z}{\mathbf{z}}
\providecommand{\uu}{\mathbf{u}}

\newcommand{\comm}[2]{\left[#1,#2\right]}
\newcommand{\acomm}[2]{\left\{#1,#2\right\}}
\newcommand{\Tr}[1]{\operatorname{Tr}\left[#1\right]}

\newcommand{\rC}{r_\text{\tiny C}}

\usepackage{hyperref}
\hypersetup{
 colorlinks=true,
 citecolor=red,
 linkcolor=blue,
 urlcolor=blue,
 pdfpagemode=UseNone,
 pdfstartview=FitH}

\begin{document}

\title{Towards relativistic generalization of collapse models}

\author{Anirudh Gundhi}
\email{anirudh.gundhi@units.it}
\affiliation{Department of Physics, University of Trieste, Strada Costiera 11, 34151 Trieste, Italy}
\affiliation{Istituto Nazionale di Fisica Nucleare, Trieste Section, Via Valerio 2, 34127 Trieste, Italy}

\author{Lajos Di\'osi}
\affiliation{Wigner Research Center for Physics, H-1525 Budapest 114 , P.O.Box 49, Hungary}
\affiliation{E\"otv\"os Lor\'and University, H-1117 Budapest, P\'azm\'any P\'eter stny. 1/A, Hungary}

\author{Matteo Carlesso}
\affiliation{Department of Physics, University of Trieste, Strada Costiera 11, 34151 Trieste, Italy}
\affiliation{Istituto Nazionale di Fisica Nucleare, Trieste Section, Via Valerio 2, 34127 Trieste, Italy}

\date{\today}
\begin{abstract}
Spontaneous collapse models provide a possible, testable solution to the quantum measurement problem. While experiments are providing increasingly stronger bounds on their parameters, a full-fledged relativistic extension is still missing. Previous attempts have encountered different obstacles, such as violation of microcausality, infinite energy rate, and particle production from vacuum. Here, we propose a generalization of the collapse master equation that is characterized by a local field collapse operator and a non-Markovian noise with a Lorentz invariant correlation. Our construction is able to overcome previously encountered problems and has the desirable properties in the non relativistic limit. A specific choice of the noise correlation function is also introduced and discussed.
\end{abstract}

\maketitle

\section{Introduction}
Non-relativistic  spontaneous wavefunction collapse (or simply, collapse) models provide a possible coherent solution to the well-known quantum measurement problem \cite{Bassi2003,Review2013}. They modify the standard, linear unitary  Schr\"odinger equation by adding non-linear and stochastic terms, which impose the statevector localization (or collapse) into one {of the} eigenstates of a suitable collapse operator. The outcome of a subsequent measurement would provide the corresponding eigenvalue. Different realizations of the stochastic process impose different outcomes, which are  distributed according to the Born rule.

While some collapse models, such as the Continuous Spontaneous Localization (CSL) \cite{PearleCSL,GhirardiCSL} and the Diosi-Penrose (DP) \cite{DiosiPenrose1987} models, are subject to  experimental and theoretical investigation \cite{Hornberger2014,bilardello2016bounds,carlesso2016experimental,vinante2017improved,torovs2018bounds,vinante2020narrowing,donadi2021underground,Carlesso2022,figurato2024effectiveness,piscicchia2024x,altamura2025improved,piccione2025exploring},  {there is still an ongoing debate concerning their possible relativistic generalizations \cite{RelCollapse1996, RelCollapse1998, RelCollapse1999,RelCollapse2004,RelCollapse2006,RelCollapse2006-b,RelCollapse2011,RelCollapse2014,RelCollapse2017,RelCollapse2019,RelCollapse2021}}.
Previous  attempts  have resulted in unwanted  effects \cite{Pearle2007,Pearle2009} such as violation of microcausality \cite{jones2021mass}, infinite energy rate   \cite{Pearle1990}, particle production from  vacuum \cite{RelCollapse2019}, and tachyon dynamics \cite{PearleTachyonic}. 
Further, a more general conceptual issue surrounds the possible compatibility between the wavefunction collapse and relativity principles \cite{Bassi2003}.

Here, we propose a mathematically consistent relativistic generalization of the {continuous collapse master equation but not any particular unravelling thereof. 
Our model} is characterized by 
a local field collapse operator and a non-Markovian stochastic  noise with a Lorentz invariant correlation. We show that 
locality saves microcausality, while non-Markovianity and a normal ordering prescription lead to a finite energy rate without  particle production from  vacuum or  tachyonic behavior. \\

\section{Non-linear unraveling}  {{Our aim is to construct the quantum field theory (QFT) version of the stochastic dynamics that resembles most closely the collapse dynamics in the non-relativistic regime at the statistical level. For that we take as starting point} the non-relativistic (NR), non-Markovian, and non-linear collapse equation for the statevector \cite{adler2007collapse,diosi2014general}. Such an equation is typically treated perturbatively in the collapse coupling constant $\gamma$ (with $\gamma$ assumed to be suitably small), although an exact, non-perturbative expression exists. For the sake of simplicity, we present the  second order expression in $\sqrt{\gamma}$,  which takes a closed form} in $\ket{\psi(t)}$ and reads \cite{adler2007collapse}
\begin{equation}\label{eq.non-linear.unrav}
    \begin{aligned}
        &\frac{\D }{\D t}\ket{\psi(t)}=\left[
       -\frac i \hbar \hat H'+\sqrt{\gamma}\sum_{i=1}^N\left(\hat A_i-\braket{\hat A_i}_t\right)\xi_i(t)\right.\\
       &+\left.\gamma \left(\hat C_+(t)-\braket{\hat C_+(t)}\right)
        \right]\ket{\psi(t)}.
    \end{aligned}
\end{equation}
 {The prescription for the higher-order equation can be found in \cite{adler2007collapse}.}
Here, $\hat H'=\hat H_0+i\hbar \gamma \hat C_-(t)$, with $\hat H_0$ being the standard quantum mechanical Hamiltonian, $\gamma$ is the common coupling with the $N$ collapse noises $\xi_i(t)$, which are real Gaussian random
processes having zero mean and correlations 
\begin{equation}
    \mathbb E \left[\xi_i(t)\xi_j(s)\right]=D_{ij}(t,s).
\end{equation}
Here $D_{ij}(t,s)$ is the to-be-determined noise correlation, $\mathbb E$ denotes the average over different realizations of $\xi_i(t)$. 
Further, $\hat A_i$ in Eq.~\eqref{eq.non-linear.unrav} are a set of commuting self-adjoint operators describing  the collapse occurrence, known as collapse operators. 
 $\hat C_-$ and $\hat C_+$ are, respectively, the anti-self-adjoint and self-adjoint parts of  $\hat C$, reading
\begin{equation}
    \hat C_\pm(t)=-\sum_{i,j=1}^N\int_0^t\D s\, D_{ij}(t,s)\comm{\hat A_i}{\hat A_j(s-t)}_\pm,
\end{equation}
where $\comm{\,\cdot\,}{\,\cdot\,}_-=\comm{\,\cdot\,}{\,\cdot\,}$ and $\comm{\,\cdot\,}{\,\cdot\,}_+=\acomm{\,\cdot\,}{\,\cdot\,}$ denote, respectively, the commutator and anticommutator, while $\hat A_j(s-t)=\hat U_0^\dag(s-t)\hat A_j\hat U_0(s-t)$  evolves freely with $\hat U_0(t)=\exp(-i \hat H_0 t/\hbar)$,  {and the initial time $t_0=0$}.  

The  {corresponding master equation (ME), which is what we aim to generalize to the relativistic regime,} can be more simply obtained  from the linear unraveling of Eq.~\eqref{eq.non-linear.unrav} for the non-normalized vector $\ket{\phi(t)}$. It reads \cite{adler2007collapse}
\begin{equation}\label{eq.linear.unrav}
\begin{aligned}
    \frac{\D}{\D t}\ket{\phi(t)}&=-\frac i\hbar {\hat H_0} \ket{\phi(t)}
  +\left[\sqrt{\gamma}\sum_{i=1}^N\hat A_i\xi_i(t)\right.\\
  &\left.-\gamma \sum_{i,j=1}^N \int_0^t\D s\,D_{ij}(t,s)\hat A_i\hat A_j(s-t)
    \right]\ket{\phi_0(t)},
\end{aligned}     
\end{equation}
where, to second order in $\sqrt{\gamma}$, we have $\ket{\phi(t)}=\ket{\phi^{(0)}(t)}+\sqrt{\gamma}\ket{\phi^{(1)}(t)}+\gamma \ket{\phi^{(2)}(t)}+\mathcal O(\gamma^{3/2})$. The corresponding ME in the Schr\"odinger picture   is $\D\hat \rho^{(2)}(t)/{\D t}=-\frac i \hbar\comm{{\hat H_0}}{\hat \rho^{(2)}(t)}+\mathcal D[\hat \rho^{(0)}(t)]$, where
\begin{equation}\label{eq.master}
    \mathcal D[\hat \rho^{(0)}(t)]\!\!=\!\!-\gamma\!\!\sum_{i,j=1}^N\!\!\int_0^t\!\!\D s\, D_{ij}(t,s)\comm{\hat A_i}{\comm{\hat A_j(s-t)}{\hat \rho^{(0)}(t)}},
\end{equation}
and $\hat \rho^{(\alpha)}(t):=\mathbb E[\ket{\phi^{(\alpha)}(t)}\bra{\phi^{(\alpha)}(t)}]$ with  $\alpha=0,1,2$. 
Moving to the interaction picture
 with $\dot{\hat \rho}^{(2)}_{I}(t)=\hat{U}_0(t)\left(\frac i \hbar\comm{{\hat H_0}}{\hat \rho^{(2)}(t)}+\dot{\hat{\rho}}^{(2)}(t)\right)\hat{U}_0^{\dagger}(t)$, we obtain
\begin{equation}
\frac{\D}{\D t}{\hat \rho}^{(2)}_\text{\tiny I}(t)\!\!=\!\!-\gamma\!\!\sum_{i,j=1}^N\!\!\int_0^t\!\!\D s\, D_{ij}(t,s)\comm{\hat A_i(t)}{\comm{\hat A_j(s)}{\hat \rho^{(0)}_\text{\tiny I}(t)}}.
\end{equation}
Now, {to generalize the dynamics to a QFT setting}, we move from the discrete label $i$ to its continuous version $\x$, and identify $\hat A_i(t)\leftrightarrow \hat Q(\x,t)$ and   
{
\begin{equation}
    \sum_{i,j=1}^N D_{ij}(t_1,t_2)\leftrightarrow \iint\D\x_1\D\x_2\,G(x_2,x_1),
\end{equation}
where $G(x_2,x_1)=\mathbb E[\xi(t_2,\x_2)\xi(t_1,\x_1)]$ and $x_1=(t_1,\x_1)$. Following this prescription, the master equation becomes
\begin{equation}
\begin{aligned}
    \frac{\D}{\D t}{\hat \rho}^{(2)}_\text{\tiny I}(t)=-\gamma\int_0^t\D t_2\iint\D\x_1\D\x_2\,G(x_2,x_1)\\\times
\comm{\hat Q(\x_1,t)}{\comm{\hat Q(\x_2,t_2)}{\hat \rho^{(0)}_\text{\tiny I}(t)}},
\end{aligned}
\end{equation}
whose solution reads
\begin{equation}
\begin{aligned}
\hat \rho^{(2)}_\text{\tiny I}(t)=\hat \rho^{(0)}_\text{\tiny I}(t)-\gamma \int_0^t\D t_1\int_0^{t_1}\D t_2\iint\D\x_1\D\x_2\,G(x_2,x_1)\\\times[\hat Q(\x_1, t_1),[\hat Q(\x_2,t_2),\hat \rho_\text{\tiny I}^{(0)}(t)]]\,.
\end{aligned}
\end{equation}}
Then, the expectation value of any generic  {local} operator $\hat{O}$  {at the generic space-time point $z = (z^0,\z)$}, to the second order in $\sqrt{\gamma}$, can be expressed as $\braket{\hat O(z)}^{(2)}=\Tr{\hat{O}_{\text{\tiny I}}(z)\hat{\rho}^{(2)}_{\text{\tiny I}}({z^0})}$. Its time derivative reads
\begin{widetext}
\begin{equation}\label{eq.mean.master}
\begin{aligned}
    \frac{\D}{\D z^0}\braket{\hat O(z)}^{(2)}&= \frac{\D}{\D z^0}\Tr{\hat O_\text{\tiny I}(z)\hat \rho(0)}-\gamma\iint_{x_2^0\leq x_1^0=z^0}\D^4x_2\,\D^3\x_1\,G(x_2,x_1)\Tr{\comm{\hat Q(x_2)}{\comm{\hat Q(x_1)}{\hat O_\text{\tiny I}(z)}}\hat \rho(0)}\\
    &-\gamma \fint\,G(x_2,x_1)\Tr{\comm{\hat Q(x_2)}{\comm{\hat Q(x_1)}{\frac{\D}{\D z^0}\hat O_\text{\tiny I}(z)}}\hat \rho(0)},
    \end{aligned}
\end{equation}
\end{widetext}
where $\fint :=\iint_{{x_2^0\leq x_1^0\leq z^0}}\D^4x_2\,\D^4x_1$.
Eq.~\eqref{eq.mean.master} can also be obtained by deriving
\begin{equation}\label{eq.solution}
\hat O^{(2)}(z)\!=\!    \hat O^{(0)}(z)\!-\!\gamma \!\fint\!\!\,G(x_2,x_1)\!\comm{\hat Q(x_2)}{\!\comm{\hat Q(x_1)}{\hat O^{(0)}(z)}},
\end{equation}
in the Heisenberg picture with respect to $z^0$, and tracing with $\hat \rho(0)$. Here, we have used $\hat O_\text{\tiny I}(z)=\hat O^{(0)}(z)$. From a mathematical standpoint,  $\hat O^{(2)}(z)$ in Eq.~\eqref{eq.solution} can also be obtained within the Heisenberg picture of a linear and unitary unraveling with $\hat{H} = \hat{H}_0+\hat H_\text{st}(t)$ where $\hat H_\text{st}(t)=\hbar\sqrt{\gamma}\int\D \z\,\hat Q(\z)\xi(t,\z)$ is a stochastic Hamiltonian. See details in Appendix \ref{App:AppendixA}. 

Conceptually, such a linear unraveling is in stark contrast with the non-linear and non-unitary unraveling of collapse models. The latter describes the objective wavefunction collapse, while the former does not. Their distinction 
is  important from foundational considerations. However, since experimentally one only has access to the density matrix \cite{Diosi_2015,Diosi2018}, here we take a more pragmatic approach and focus only on the latter dynamics, namely the ME. {For this reason, in what follows, we will work with the unitary unraveling in the Heisenberg picture. This has the added advantage that the time evolution of the operators can be obtained to any desired order in $\sqrt{\gamma}$ by using standard perturbation theory. See details in Appendix \ref{App:AppendixA}.}

Note that  $\hat{Q}(x)$  in Eq.~\eqref{eq.solution} is the freely evolved operator in the Heisenberg picture and $G(x,y)$ some generic function of $x$ and $y$. Next we study the requirements on $\hat{Q}(x)$ and $G(x,y)$ so  that Eq.~\eqref{eq.solution} is relativistically consistent.\\

\section{Lorentz covariance} The Lorentz covariance of Eq.~\eqref{eq.solution}  {can be ensured via} standard QFT {considerations}. For the operator $\hat O(z)$ to evolve covariantly, it is sufficient to show that the  {collapse} noise does not introduce any additional transformation for  $\hat O$ in a reference frame transformation, and that $\hat O$ transforms  the same  as $\hat O^{(0)}$. The first term on the RHS of Eq.~\eqref{eq.solution} trivially satisfies this requirement. Further, assuming that $\hat Q(x)$ is  a Lorentz scalar, and the correlation  $G(x,y)$ to be  Lorentz invariant   $G(x,y)=G\left((x-y)_{\mu}(x-y)^{\mu}\right)$, we see that the second term of Eq.~\eqref{eq.solution} also transforms covariantly. Thus, Lorentz covariance is guaranteed as long as the collapse operator $\hat{Q}(x)$ is a Lorentz scalar and $G(x-y)$ a Lorentz invariant function.  {Note that Lorentz covariance holds at all orders, and not only to second order in $\sqrt{\gamma}$. This is shown below Eq.~\eqref{eq:Pert_exact} in the Appendix  \ref{App:AppendixA}.}\\

\section{The microcausality condition (MCC)}  
The MCC states that
\begin{equation}\label{eq.micro}
    \comm{\hat{O}(z_2)}{\hat{O}(z_1)}=0,\quad\text{for}\quad |z_2-z_1|<0,
\end{equation}
 where we use the {$(+,-,-,-)$} convention.  MCC implies that the measurement of one observable cannot influence the time evolution of any other observable outside of the lightcone corresponding to the measurement event. This request is met within standard QFT for typical interactions \cite{Weinberg1995,Peskin1995}. 
We now study MCC for operators evolving according to a unitary unraveling of Eq.~\eqref{eq.solution}, where the  evolution of the operators is governed by the Hamiltonian $\hat{H}(t)= \hat{H}_0+\hat{H}_{\text{st}}(t)$. Namely
\begin{equation}\label{eq.U.unitary}
\hat{U}_{t;t_0} = \mathcal{T}\left\lbrace\exp{\left(-\frac i\hbar\int_{t_0}^t\D t'\, \hat{H}(t')\right)}\right\rbrace.
\end{equation}
{Note that the evolution of the statevector can be written in a manifestly Lorentz covariant manner within the Tomonaga-Schwinger formalism; the corresponding evolution then reads $\ket{\psi_\sigma}=\hat U_{\sigma;\sigma_0}\ket{\psi_{\sigma_0}}$, where $\sigma_0$ and $\sigma$ are the initial and final spacelike hypersurfaces respectively and $\hat U_{\sigma;\sigma_0}=\mathcal{T}\left\lbrace\exp{\left(-\frac i\hbar\int_{\sigma_0}^\sigma\D^4 z'\, \hat{\mathcal{H}}_\text{st}(z')\right)}\right\rbrace$ \cite{Bassi2003}. Since we demonstrated the Lorentz covariance without resorting  to the Tomonaga-Schwinger formalism, for the sake of simplicity, we continue the analysis in a given reference frame.}

The unitary operator in Eq.~\eqref{eq.U.unitary} satisfies $\hat U_{t_2;t_1}\hat U_{t_1;t_0}=\hat U_{t_2;t_0}$ for any Hermitian operator $\hat H(t)$. In the Heisenberg picture where $\hat{O}(t, \z) =   \hat{U}^{\dagger}_{t;t_0}\hat{O}(t_0,\z)\hat{U}_{t;t_0}$, the MCC condition, for $|z_2-z_1|<0$, requires
\begin{equation}\label{eq:MCC_Unit}
    \hat U_{t_1;t_0}^\dag\comm{\hat U_{t_2;t_1}^\dag\hat O(t_0,\z_2)\hat U_{t_2;t_1}}{\hat O(t_0,\z_1)}\hat U_{t_1;t_0}=0.
\end{equation}
In general, $\hat U_{t_2;t_1}$ does not satisfy the time-translation property --- $\hat U_{t_2;t_1}= \hat U_{\Delta t+t_0;t_0}$, where $\Delta t=t_2-t_1$, if the Hamiltonian $\hat H(t)$ is time-dependent. However, since the time dependence appears in $\hat H_\text{st}(t)={\hbar\sqrt{\gamma}}\int\D\z\,\hat Q(\z)\xi(t,\z)$ only through the noise, we can rewrite the time integral in Eq.~\eqref{eq.U.unitary} as
\begin{equation}
    \hat Q(\z)\int_{t_1}^{t_2}\D t'\,\xi(t',\z)=    \hat Q(\z)\int^{\Delta t+t_0}_{t_0}\D s\,\tilde\xi(s,\z).
\end{equation}
Therefore, $\hat U^{(\xi)}_{t_2;t_1}$ encoding the stochastic dynamics in our analysis can be written as $\hat U_{t_2;t_1}^{(\xi)}=\hat U_{\Delta t+t_0;t_0}^{(\tilde{\xi})}$.
We point out that $\tilde\xi(t,\z)=\xi({t+t_1-t_0},\z)$ has the same correlation function, due to the invariance of $G$ under  {spacetime} translations. Therefore, at the level of the ME, the distinction between $\xi$ and $\tilde{\xi}$ is unimportant and will not be retained in what follows.

\begin{figure}[t]
	\centering
	\includegraphics[width=\linewidth]{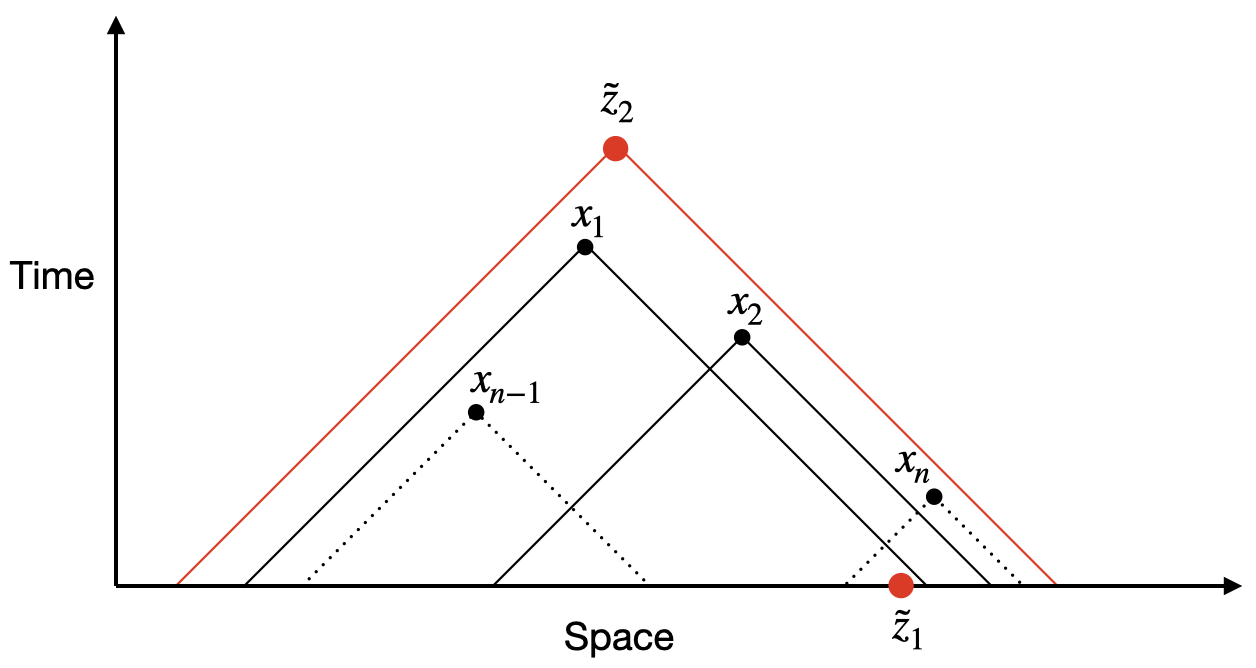}
	\caption{Leading order causal structure which demonstrates why MCC is respected, cf.~Eq.~\eqref{eq.DoubleComMicro} and Eq.~\eqref{eq.DoubleComMicro1}. }\label{Im:Lightcones}
\end{figure}

{In this context}, since $\hat U_{t_2;t_1}$ is equivalent to $\hat U_{\Delta t+t_0;t_0}$, MCC specified in Eq.~\eqref{eq:MCC_Unit} is equivalent to showing that for $|z_2-z_1|<0$, $\comm{\hat O(\Delta t+t_0,\z_2)}{\hat O(t_0,\z_1)}=0$. Note that now only the first operator within the commutator evolves according to the collapse dynamics. To second order in $\sqrt{\gamma}$, for a given realization of the noise, the latter expression reads
\begin{equation}\label{eq.DoubleComMicro}
\fint \xi(x_2)\xi(x_1)\comm{{\comm{\hat Q(x_2)}{\comm{\hat Q(x_1)}{\hat {O}^{(0)}(\tilde z_2)}}}}{\hat {O}^{(0)}(\tilde z_1)},
\end{equation}
where $\tilde z_2=(t_0+\Delta t,\z_2)$ and $\tilde z_1=(t_0,\z_1)$.
The integral $\fint$ implies that $x_2^0\leq x_1^0$ and $x_1^0\leq\tilde{z}^0_2$, where $\tilde{z}_1^0$ sets the initial time.
For  Eq.~\eqref{eq.DoubleComMicro} to be non-zero, its innermost  commutator $[\hat Q(x_1),\hat{O}^{(0)}(\tilde{z}_2)]$ must also be  non-zero. Since both the operators appearing in such a commutator are local QFT operators evolving with respect to the standard free Hamiltonian $\hat H_0$,  $x_1$ must be inside the past lightcone of $\tilde{z}_2$.  {Note that} $x_1$ is guaranteed not to be in the future of $\tilde{z}_2$ by the standard perturbative expansion. Similarly, for the double commutator involving $\hat{Q}(x_2)$, $\hat{Q}(x_1)$ and $\hat{O}^{(0)}(\tilde{z}_2)$ to be non-zero, $x_2$ must belong to the past lightcone of $\tilde{z}_2$ or $x_1$, which makes it necessary for $x_2$ to also belong to the past lightcone of $\tilde{z}_2$ as shown in Fig.~\ref{Im:Lightcones}. 
Finally, by following the same logic, for the outermost commutator involving $\hat{O}^{(0)}(\tilde{z}_1)$  to be non-zero, the necessary condition is that $z_1$ must belong to the past lightcone of $\tilde{z}_2$ or $x_1$ or $x_2$, which forces it inside the past lightcone of $\tilde{z}_2$. Thus, MCC in Eq.~\eqref{eq.micro} is  respected to second order due to the locality of the operator $\hat{Q}(x)$ as in standard QFT.

Following the same reasoning, it can  be shown that MCC is respected to all orders in $\sqrt{\gamma}$. Indeed, in the  $n$-th order term
\begin{equation}\label{eq.DoubleComMicro1}
\fint\! \xi(x_n)...\xi(x_1)\comm{{\comm{\hat Q(x_n),...}{\comm{\hat Q(x_1)}{\hat {O}^{(0)}(\tilde{z}_2)}}}}{\hat {O}^{(0)}(\tilde{z}_1)},
\end{equation}
$\tilde{z}_1$ must lie inside the past lightcone of $\tilde{z}_2$ or at least one of the coordinates $x_1, x_2,...,x_n$ of the collapse operators. Since $x_1, x_2,...,x_n$ themselves must lie in the past lightcone of $\tilde{z}_2$, it becomes necessary for $\tilde{z}_1$ to lie inside the past lightcone of $\tilde{z}_2$ for  $[\hat{O}(\tilde{z}_2),\hat{O}(\tilde{z}_1)]$ to be non-zero, and therefore ${z}_1$ to lie inside the past lightcone of ${z}_2$ for  $[\hat{O}({z}_2),\hat{O}({z}_1)]$ to be non-zero (having assumed $t_2\geq t_1$ without loss of generality). Therefore, the locality of $\hat{Q}$ and the standard time-ordered evolution, which imposes the past lightcone structure in Fig.~\ref{Im:Lightcones},  together ensure that MCC is respected for the relativistic stochastic dynamics as well. This proof shows that if $\hat{Q}$ is a non-local Lorentz scalar, or if the time ordering is removed, then MCC might be violated.
An example of the former is the non-local collapse operator  $\hat Q\propto\hat{\varphi}_{+}\hat{\varphi}_{-}$ proposed in \cite{jones2021mass}, where $\hat{\varphi}_{+}$ and $\hat{\varphi}_{-}$ are respectively the positive and negative frequency parts of $\hat{\varphi}$. Instead, an example of the latter is \cite{PearleTachyonic} where MCC was  violated  due to the removal of time-ordering being imposed to remove divergences. Our analysis  shows that the noise correlations do not need to be ultra-local ones, such as $G(x,y)=\delta^4(x-y)$, to preserve MCC as argued in \cite{jones2021mass}. Instead, MCC can be satisfied by taking a local collapse operator and non-divergent correlations.  
 
 In our work, motivated by the NR CSL model, we choose the collapse operator to be 
\begin{equation}\label{eq:ColOp}
\hat Q(x)=\frac{1}{2}\alpha\hat \varphi^2(x),
\end{equation}
where $\alpha$ is a suitable free parameter. $\hat{Q}(x)$ is local, and it becomes proportional  to the mass density operator $\hat Q_\text{\tiny MD}(x)$ in the NR limit. We emphasize that the
choice of $\hat Q$ is not unique. One might also consider $\hat Q$ constructed from the energy-momentum tensor $\hat T_{\mu \nu}$.
 As argued before, requiring  $\hat Q$ to be a Lorentz scalar ensures the Lorentz covariance of the collapse dynamics, for example one might also consider $\hat Q\propto\hat{T}^{\mu}_{\mu}$. While details might change, we expect our prescription to hold as long we consider $\hat Q$ being local and
quadratic in creation and annihilation operators.

\section{Energy rate} Now we show that a finite energy rate can be achieved as long as $G(x,y)$ is well-behaved. Setting $\hbar=c=1$, we compute the energy rate $\D E/\D z^0$ with Eq.~\eqref{eq.mean.master} by taking $\hat O(z)=\hat{\mathcal H} (z)$, where
\begin{equation}
    \hat{\mathcal H}(z)=\tfrac12\hat \pi^2(z)+\tfrac12 \left(\nabla\hat \varphi(z)\right)^2+\tfrac12m^2\hat \varphi^2(z),
\end{equation}
and then integrating over space, i.e.~$\z$.
In doing so, the first and  last term on the RHS of Eq.~\eqref{eq.mean.master} vanish, as  the freely evolved Hamiltonian is a conserved quantity with $\D/\D z^0  \int\D\z\,\hat{\mathcal H}^{(0)}(z)=0$. Therefore, we have 
\begin{equation}\label{eq.E1}
\begin{aligned}
    \frac{\D E}{\D z^0}= -\gamma\iiint_{x^0\leq y^0=z^0}\D^4x\,\D^3\y\,\D^3\z\,G(x,y)\\\times\Tr{\comm{\hat Q(x)}{\comm{\hat Q(y)}{\hat{\mathcal H}(z)}}\hat \rho(0)}.
    \end{aligned}
\end{equation}
Since $y^0=z^0$, the inner commutator can be computed in the standard way and reads
\begin{equation}
    \comm{\hat Q(y)}{  \hat{\mathcal H}(z)}=\frac{i\alpha}{2}   \delta^3(\y-\z)\left\lbrace\hat\varphi(z), \hat\pi(z)\right\rbrace,
\end{equation}
giving
\begin{equation}   \label{eq.defND}
    \begin{aligned}
    \frac{\D E}{\D z^0}= -{\gamma\alpha^2}\iint_{x^0\leq z^0}\D^4x\,\D^3\z\,G(x,z)\\\times \partial_{z^0}\left(\mathcal N(x,z)\mathcal D(x-z)\right),
    \end{aligned}
\end{equation}
where 
\begin{equation}\label{eq.E2}
    \begin{aligned}
        \mathcal N(x,z)&:=\frac12 \Tr{\acomm{\hat \varphi(x)}{\hat \varphi(z)}\hat \rho(0)},\\
        \mathcal D(x-z)&:=i\comm{\hat \varphi(x)}{\hat \varphi (z)}.
    \end{aligned}
\end{equation}
Here, we have used the fact that $\partial \hat\varphi(z)/\partial z^0=\hat \pi(z)$. The expression for energy rate simplifies further, if the initial state is such that the LHS of the first line in Eq.~\eqref{eq.E2} is a function of $x-z$ (for example, a thermal state).  Since Lorentz covariance implies $G(x,z)=G(x-z)$, Eq.~\eqref{eq.defND} would only depend on the spatial and temporal parts $\uu=\x-\z$ and $\tau=x^0-z^0$ respectively, such that
\begin{equation}\label{eq.E3}
    \frac{\D E}{\D z^0}=\gamma\alpha^2 V\int\D^3  \uu\int_{-\infty}^0\D\tau\,G(\tau,\uu){\partial_{\tau}}\left(\mathcal N(\tau,\uu)\mathcal D(\tau,\uu)\right),
\end{equation}
where $V=\int\D\mathbf{v}$ is the full volume, with $\mathbf{v}:=(\x+\z)/2$.\\

\section{The non-Markovian model} 
To arrive at a compact expression for the rate of  energy for a general non-Markovian noise, we assume that the initial state $\hat\rho(0)$ has a definite particle number (like the thermal state). Then, we have
\begin{equation}\label{eq.InitialState}
\Tr{\hat\rho(0) \hat{a}_{\q}\hat{a}_{\p}} = \Tr{\hat\rho(0) \hat{a}^{\dagger}_{\q}\hat{a}^{\dagger}_{\p}}=0,
\end{equation}
where $\hat{\varphi}(x) := \hat{\varphi}_{+}(x)+\hat{\varphi}_{-}(x)$, with
\begin{equation}
\hat{\varphi}_{+}(x):=\int\frac{\D^3\p}{\sqrt{2\omega_p(2\pi)^3}}\,e^{-ip.x}\hat a_\p,\qquad p=(\omega_p,\p).
\end{equation} 
Even though such an initial state is not Lorentz invariant, it serves the  purpose of demonstrating that a Lorentz invariant non-Markovian noise is free of the  problems of its Markovian counterpart,  which is well-known to have a  $\delta(\mathbf{0})$ divergence, independently of the choice of the initial state (c.f.~Appendix \ref{App:AppendixB}). 

The application of Eq.~\eqref{eq.InitialState} to Eq.~\eqref{eq.E2} provides the following simple expressions:
\begin{equation}\label{eq.defND2}
    \begin{aligned}
        \mathcal N(x)&=\int\frac{\D^3\q}{(2\pi)^3}\left(n_{\q}+\frac12\right)\frac{\cos(q.x)}{\omega_q},\\
        \mathcal D(x)&=\int\frac{\D^3\q}{\omega_q(2\pi)^3}\sin(q.x),
    \end{aligned}
\end{equation}
where $n_\q$ is the average occupancy of the mode $\q$ for the initial state, such that $\int \D^3\q \,n_{\q}\to \frac{(2\pi)^{3/2}}{L^3}\sum n_{q}= (2\pi)^{3/2}N/V$, $N$ being the total number of particles. The factor of $1/2$ in the first expression captures the standard QFT vacuum  divergence.  
This factor leads to a divergent  particle production rate from vacuum, which was already found in Ref.~\cite{PearleTachyonic}. There, the removal of time-ordering was proposed to obtain a finite expression. However, as argued before, this might violate causality, which is also reflected in the tachyonic behavior reported in Ref.~\cite{PearleTachyonic}. Therefore, here, we propose to use a normal-ordering prescription, where all the observables of interest are normal-ordered at all times, i.e.~$\hat{O}(z)\rightarrow :\hat{O}(z):$ and thus  the $1/2$ term  drops. Note that this prescription leaves the MCC analysis unchanged, as shown in Appendix \ref{App:AppendixC}. 

By combining Eqs.~\eqref{eq.defND2} and~\eqref{eq.E3}, and computing explicitly the $z^{0}$ derivative, we can derive the following  normal-ordered (NO)  expression
\begin{equation}\label{eq.NO1}
\begin{aligned}
        \left.\frac{\D E}{\D z^0}\right|_{\text{NO}}&=\frac{\gamma \alpha^2 V}{(2\pi)^6}\int\D^3\x\int_{-\infty}^0\D\tau\, G(\tau,\x)
        \iint\D^3\p\,\D^3\q\, n_{\q}\\
        &\times\left[\frac{\cos(p.x)\cos(q.x)}{\omega_q}-\frac{\sin(p.x)\sin(q.x)}{\omega_p}\right].
\end{aligned}
\end{equation}
Since its integrand  remains unchanged under reflection $x\to-x$, Eq.~\eqref{eq.NO1} can be written more compactly as 
\begin{equation}\label{eq.relDERN}
\begin{aligned}
        &\left.\frac{\D E}{\D z^0}\right|_{\text{NO}}=\frac{\gamma \alpha^2 V}{2(2\pi)^4}\iint\D^3\p\,\D^3\q\, n_{\q} \\
        &\times\Re\left[\frac{\mathcal G(p+q)+\mathcal G(p-q)}{2\omega_q}+\frac{\mathcal G(p+q)-\mathcal G(p-q)}{2\omega_p}\right],
\end{aligned}
\end{equation}
where $\mathcal G(p)$ is the four-dimensional Fourier transform of the correlation  $G(x)$.  Notably, the multiplicative $\delta(\mathbf{0})$ divergence present the Markovian noise expression does not appear in the non-Markovian case. Further, as long as one is only concerned with obtaining a finite  energy rate, any choice of $\mathcal{G}(q)$ for which  Eq.~\eqref{eq.relDERN} is finite becomes a viable  relativistic non-Markovian collapse correlation  within the normal-ordering prescription.\\

\section{Non-relativistic and relativistic limits}
We now discuss the NR limit to motivate specific choices  for  $\mathcal{G}(q)$ {which, for it to be Lorentz inavariant, should be a function of $q^2$, i.e.~$\mathcal G(q)\to\mathcal G(q^2)$}. 
Such a limit can be understood as that for which $n_{\q}=0$, for $\q^2\gtrsim m^2$. Namely, only particles whose kinetic energy is well below the rest mass energy are retained. This also implies $\omega_q\approx m$. Notably, the $\p$ integral in Eq.~\eqref{eq.relDERN} still runs over all the $\mathbb R^3$ values. 
We divide the $\p$ integral into the NR ($|\p|^2\ll m^2$ and $\omega_p\approx m$) and the relativistic ($|\p|^2\gg m^2$ and $\omega_p\approx |\p|$) regimes. 
Since the noise correlation is Lorentz invariant, with the following structure $\mathcal{G}(p\pm q) \to \mathcal{G}\left[(\omega_{p}\pm\omega_{q})^2- (\p\pm\q)^2\right]$, 
in the NR regime we have   ${\mathcal{G}_{\mathrm{NR}}(p+q)\to \mathcal{G}(4m^2)}$ and ${\mathcal{G}_{\mathrm{NR}}(p-q)}/\omega_q-{\mathcal{G}_{\mathrm{NR}}(p-q)}/\omega_p\approx0$.
On the other hand, when $\p$ is in the relativistic regime, we have $\mathcal{G}_{\mathrm{rel}}(p\pm q)\to \mathcal{G}(\pm 2|\p|m)$. Since $\mathcal{G}(q)$ is an even function --- which follows from the requirement that ${G}(x)$ must be real --- $\mathcal{G}_{\mathrm{rel}}(p+q)- \mathcal{G}_{\mathrm{rel}}(p-q)\approx 0$, and the rate  of increase of energy  becomes
\begin{equation}
\begin{aligned}\label{eq.Nrel_DE_RN}
\left.\frac{\D E}{\D z^0}\right|_{\text{NO}}&\approx\frac{\gamma \alpha^2 V}{2(2\pi)^4}\int_{0}^{m}\D\q\, n_{\q}\\
&\times\left[\int_{0}^{m}\D\p\,\frac{\mathcal G(4m^2)}{m}+\int_{m}^{\infty}\D\p\,\frac{\mathcal G(2|\p|m)}{m}\right],
\end{aligned}
\end{equation}
where the first integral in brakets gives $\mathcal G(4m^2)/m \times 4/3 \pi m^3$. Thus, in the NR limit, we find that the energy rate is proportional to the total number of particles $N$, as the overall factor of $V$ in Eq.~\eqref{eq.Nrel_DE_RN} cancels out due to the relation $\int \D\q\, n_{\q}=(2\pi)^{3/2}N/V$. Such a feature is shared with the NR  collapse models in literature \cite{Carlesso2022}. Further, the choice of $\mathcal{G}(q)$ is now only constrained by the convergence of the second integral in Eq.~\eqref{eq.Nrel_DE_RN}, {which can be easily achieved by a suitable choice of $\mathcal{G}(q)$.
Nevertheless, for the relativistic Markovian  noise $G(x)=\delta^4(x)$,  $\mathcal G(q)$ is a constant and we get that Eq.~\eqref{eq.Nrel_DE_RN} diverges as $\delta(\textbf{0})$, as expected and also shown in Appendix \ref{App:AppendixB}. }

In the fully relativistic limit, for which the corresponding distribution $n(\q)$  is non-zero only when $|\q|\gtrsim m$, we can find a finite expression for the energy rate.
By following a similar reasoning as before, Eq.~\eqref{eq.relDERN} can be approximated to
\begin{equation}\label{eq.Rel_DE_RN}
\begin{aligned}
&\left.\frac{\D E}{\D z^0}\right|_{\text{NO}}\approx\frac{ \gamma \alpha^2 V}{2(2\pi)^4}\int_{m}^{\infty}\D\q\, \frac{n_{\q}}{|\q|}\\
&\times\left[\int_{0}^{m}\D\p\,{\mathcal G(2m|\q|)}+\int_{m}^{\infty}\D\p\,\mathcal G\left(4|\p||\q|\sin^2{\theta_{pq}}\right)\right],
\end{aligned}
\end{equation}
where $\theta_{pq}$ is the angle between the four-dimensional vectors $p$ and $q$, and we can substitute the first integral in brakets with $\mathcal G(2m|\q|)\times 4/3\pi m^3$. The convergence of Eq.~\eqref{eq.Rel_DE_RN} puts rather mild constraints on $\mathcal{G}(q)$, and thus on $G(x)$.\\

\section{Outlook} {To complete  the model we need to specify $\mathcal G(q^2)$. The natural choice would have been $\mathcal G(q^2)\propto \exp({-a^2q^2})$ as in the NR CSL model. However, this is not suitable since $q^2$ becomes negative for the spacelike regions and $\mathcal G(q)$ diverges. We therefore propose 
\begin{equation}\label{eq.NoiseKernel}
\mathcal{G}(q) \to \mathcal G(q^2)= \exp{\left(-(q^2)^2/\beta^4\right)},
\end{equation}
where $\beta$ is another free parameter of the model. This choice is not unique; even in the NR collapse models the spatial correlation of the noise is not determined from first principles. This is also reflected in the relativistic case, where other options, e.g.~$\theta(-q^2)\exp(a^2q^2)$, have been already considered in \cite{Pearle2009}. Our choice in Eq.~\eqref{eq.NoiseKernel} is motivated for the sake of simplicity while remaining close to the choice in the NR CSL model. In particular,}
%
%
$\mathcal{G}(q^2)$ in Eq.~\eqref{eq.NoiseKernel} is finite in any point in the Fourier space; $\mathcal G(q^2)\to 0$ fast enough as $q^2\to \pm \infty$ thereby guaranteeing the convergence of Eq.~\eqref{eq.relDERN}. Moreover, in $(1+1)$ dimension, one can also obtain an analytic expression for the corresponding ${G}(x)$, which is given in terms of the Meijer G-function
\begin{equation}\label{eq.Noise}
G(x) = \frac{\beta^2}{2}G^{2,0}_{0,3}\left(\left.\frac{1}{256}\beta^4(x^2-t^2)^2\right|0,0,\frac{1}{2}\right).
\end{equation}
Such an expression features a peak at a characteristic length scale $x=t$ (with  $c=1$) and goes to zero for large spatial separations. While an analytic expression for $G(x)$ in $(1+3)$ dimensions is hard to obtain, it is reasonable to assume that $G(x)$  would still have the desired properties as its $(1+1)$ counterpart. 

Along with the collapse operator choice, the choice of $G(x)$ [cf.~Eq.~\eqref{eq.Noise}] completely fixes the relativistic stochastic dynamics up to the free parameters $\alpha$, $\beta$ and $\gamma$. 
These parameters,  like in the CSL model, will be subject to experimental investigation. {For instance, already in the NR limit, Eq.~\eqref{eq.Nrel_DE_RN} gives (as a rough estimate) $\D E/\D z^0\propto \gamma \alpha^2 m^2 N \exp(-16 m^4/\beta^4)$. Then, the free parameters of our model  are subject to experimental bounds given by $\D E/\D z^0/(m N)\simeq 2\times 10^{-11}\,$W/kg \cite{Carlesso2022}.
Note that our expression  has a different mass dependence compared to the corresponding NR CSL expression $\D E/\D t\propto \gamma m N/\rC^5$ \cite{Carlesso2022}, where $\rC$ is the spatial correlation length of the CSL noise. 
}

\section{Discussion} We have shown that a consistent relativistic stochastic dynamics  can be constructed, which leads to a finite energy rate and also respects the microcausality condition. For a local quadratic  field collapse operator, a collapse noise with a non-Markovian Lorentz invariant correlation, and with an additional normal ordering prescription, such a dynamics is free of the problems found in previous works and has a desirable CSL like behavior in the NR limit.  

However, even after such technical difficulties have been overcome, the question of whether such a dynamics can be consistently interpreted as a relativistic collapse model is still open. It has been shown in Sec.~14.2 of \cite{Bassi2003} that even though a consistent relativistic stochastic dynamics might be constructed at the level of the density matrix (obtained after averaging over different realizations of the collapse noise), for a single specific realization of the collapse noise, the quantum expectation value can in general be different when computed along different hypersurfaces by  observers moving relative to each other. This poses a conceptual problem if one wishes to assign an objective meaning to the statevector at all times for each realization of the stochastic noise. One might not find such a difficulty surprising, given that related conceptual issues concerning {non-Markovian} quantum monitoring have also been pointed in  \cite{Diosi1990RelQF,diosi2008non,diosi2008nonErratum,wiseman2008pure}. {Nevertheless, a resolution to such a conceptual issue has been proposed in \cite{RelCollapse2014} for multiple local measurements. We emphasize that independently of these important discussions concerning the evolution of the wavefunction, our work still offers a possible consistent relativistic non-Markovian dynamics via the master equation.}

{In this view, it would be interesting to explore possible technical and conceptual overlap of our work with those in the framework relativistic open quantum system dynamics \cite{breuer2002theory,calzetta2009nonequilibrium,fogedby2022field,kashiwagi2025effective,bowen2025open}. Such a connection is not obvious. For instance, in the open quantum system framework, one needs to make an additional  choice, being that of the total system-environment initial state when tracing over the environmental fields. Moreover, 
 typical choices for the initial state of the environment (apart from the vacuum state) break the Lorentz covariance of the system master equation. 
 
Conversely, in the framework we describe here, the environmental fields are substituted with noise fields, which are defined in terms of two-point correlations that are not necessarily associated to an initial QFT state. This makes the
associated dynamics  less restrictive compared to that of relativistic open quantum systems, by allowing to choose viable two-point correlations [cf.~Eq.~\eqref{eq.NoiseKernel}] to construct a Lorentz covariant master equation.
Nevertheless, in future works, such a connection would be useful to build, allowing to exploit  well-established QFT techniques, for example, those dealing with renormalization \cite{agon2018divergences,nagy2022renormalizing}.} \\

\section*{Acknowledgments} The authors thank Lorenzo Di Pietro and Angelo Bassi for several discussions. The authors acknowledge support from the Department of Physics of the University of Trieste (Department of Excellence 2023-2027), University of Trieste, INFN, the PNRR MUR projects PE0000023-NQSTI, the EU EIC Pathfinder project QuCoM (101046973), the FVG MICROGRANT LR2/2011 (D55-microgrants24)    , the National Research, Development and Innovation Office “Frontline” Research Excellence Program (Grant No. KKP133827) and  the EU COST Actions CA23115 and CA23130.

\bibliography{main}{}
\appendix

\section{The unitary unraveling} \label{App:AppendixA}
We consider the quantum dynamics governed by the Hamiltonian  $\hat{H}$, 
\begin{equation}
\hat{H}=\hat{H}_{0}+ \hat{H}_{\text{st}}(t),
\end{equation}
where $\hat{H}_{0}$ is the standard Hamiltonian of the real free Klein-Gordon scalar field, while $\hat{H}_{\text{st}}(t) = \hbar\sqrt{\gamma}\int\D \z\,\hat Q(\z)\xi(t,\z)$ is the contribution from the collapse noise within the unitary unraveling.  The effect of the latter can be studied perturbatively, as in standard quantum field theory (QFT). For that, as it is well-known, the full unitary operator $\hat{U}_{t;t_0}$ can be factorized as
\begin{equation}\label{aeq:UnitaryFact}
\begin{aligned}
    \hat{U}_{t;t_0} &= \hat{U}_{0}(t,t_0)\times \hat{U}_{\text{st}}(t,t_0),\\ \hat{U}_{0}(t,t_0)&=\exp{\left[-\frac{i}{\hbar}\hat{H}_{0}(t-t_0)\right]},\\ \hat{U}_{\text{st}}(t,t_0)&=\mathcal{T} \left\{\exp \left[-\frac{i}{\hbar}\int_{t_0}^t \D t'\hat{H}_{\text{st}}^\text{\tiny I} (t') \right]\right\},
\end{aligned}
\end{equation}
where $\hat{H}_{\text{st}}^\text{\tiny I} (t) = \hat{U}^{\dagger}_0(t,t_0)\hat{H}_{\text{st}}(t)\hat{U}_0(t,t_0)$. In the Heisenberg picture, given Eq.~\eqref{aeq:UnitaryFact}, a generic operator $\hat{O}(z)$, to second order, is given by
\begin{align}
&\hat{O}(\z,t)=\nonumber\\
&\left[\hat{1}+\frac{i}{\hbar}\int_{t_0}^t \D t' \hat{H}_{st}^\text{\tiny I} (t')  - \frac{1}{\hbar^2}\int_{t_0}^t \D t' \int_{t_0}^{t'}  \D t'' {\hat{H}_{\text{st}}^\text{\tiny I}(t'') \hat{H}_{\text{st}}^\text{\tiny I}(t')}\right]\nonumber\\&\times\hat{O}_0(\z,t)\times\nonumber\\
&\left[\hat{1}-\frac{i}{\hbar}\int_{t_0}^t \D t' \hat{H}_{\text{st}}^\text{\tiny I} (t')-\frac{1}{\hbar^2}\int_{t_0}^t  \D t'\int_{t_0}^{t'} \D t'' \hat{H}_{st}^\text{\tiny I}(t') \hat{H}_{\text{st}}^\text{\tiny I} (t'')\right]\nonumber\\
&+\text{higher order}.
\end{align}
The time evolution of $\hat{O}$ can then be written in terms of the  freely evolved ($0^{\text{th}}$ order) term   $\hat{O}^{(0)}(t) = \hat{O}_0(t) = \hat{U}^{\dagger}_0(t,t_0)\hat{O}\hat{U}_0(t,t_0)$ as
\begin{align}
\hat{O}(\z,t) &=\hat{O}^{(0)}(\z,t)+\frac{i}{\hbar} \int_{t_0}^{t} \D t' \left[\hat{H}_{\text{st}}^\text{\tiny I} (t') , \hat{O}^{(0)}(\z,t)\right]\nonumber\\
&-\frac{1}{\hbar^2}\int_{t_0}^t  \D t'\int_{t_0}^{t'} \D t'' {\left[\hat{H}_{\text{st}}^\text{\tiny I}(t''),\left[\hat{H}_{\text{st}}^\text{\tiny I}(t'),\hat{O}^{(0)}(\z,t) \right] \right]}\nonumber\\
&+\text{higher order}\,.
\end{align}
Substituting the expression for $\hat{H}_{\text{st}}$ and averaging  over the stochastic realizations of the noise $\xi(x)$, we get
\begin{align}\label{aeq.solution}
\mathbb{E}[\hat O^{(2)}(z)] =&\,\, \hat O^{(0)}(z)-\gamma\iint_{x^0\leq y^0\leq z^0}\D^4x\,\D^4y\nonumber\\
&\times\mathbb{E}[\xi(x)\xi(y)]\,\comm{\hat Q(x)}{\comm{\hat Q(y)}{\hat O^{(0)}(z)}}\nonumber\\
=&\,\,\hat O^{(0)}(z)-\gamma\iint_{x^0\leq y^0\leq z^0}\D^4x\,\D^4y\nonumber\\
&\times G(x-y)\comm{\hat Q(x)}{\comm{\hat Q(y)}{\hat O^{(0)}(z)}},
\end{align}
whose derivative with respect to $z^0$ gives Eq.~\eqref{eq.mean.master} of the main text, {which was derived starting from the non-unitary unraveling of the statevector in Ref.~\cite{adler2007collapse}. For the unitary unraveling within the Heisenberg picture, noise-averaged time evolution of any operator can be evaluated to higher orders using standard perturbation theory. To an arbitrary order $2n$ it reads
\begin{equation}\label{eq:Pert_exact}
\begin{aligned}
&\mathbb{E}[\hat O^{(2n)}(z)] = \hat O^{(0)}+\dots+\left(i\sqrt{\gamma}\right)^{2n}\times\\
&\fint\mathbb{E}\left[\xi(x_{2n})\dots\xi(x_1)\right]\comm{\hat Q(x_{2n}),\dots}{\comm{\hat Q(x_1)}{\hat {O}^{(0)}(z)}}.
\end{aligned}
\end{equation}
Just as Lorentz covariance is respected to second order in $\sqrt{\gamma}$, it can be seen that if $\hat{Q}$ is a Lorentz scalar and the two-point correlation $\mathbb{E}\left[\xi(x_1)\xi(x_2)\right] = G(x_1,x_2)$ is Lorentz invariant, then Lorentz covariance holds to all orders in $\sqrt{\gamma}$. This is because, as with the second order term, the higher order terms do not change the standard QFT transformation of the operator $\hat O^{(2n)}$. That is,  $\hat O^{(2n)}$ still transforms as $\hat O^{(0)}$. This can be seen from the fact that all the operators $\hat{Q}(x_{2n}),\dots,\hat Q(x_1)$ are Lorentz scalars and that for the Gaussian noise (with non-white correlation) the expectation value $\mathbb{E}\left[\xi(x_{2n})...\xi(x_1)\right]$ can be written as a product of two-point Lorentz covariant correlations, summed over all possible contractions, and is therefore Lorentz invariant. Thus, the operator $\hat{O}$ would transform in the same way in the presence of the collapse noise, as it does in standard QFT, to all orders in $\sqrt{\gamma}$.} 

\section{The Markovian noise} \label{App:AppendixB}
The request of having a Lorentz invariant noise correlation together with the request of Markovian noise, i.e. $\mathbb{E}[\xi(x^{0},\x)\xi(z^{0},\z)]= \delta(\tau)f(\uu)$, $\tau=x^{0}-z^{0}, \uu=\x-\z$, fixes the noise correlation to  $G(\tau,\uu) = \delta(\tau)\delta(\uu)$. For $G(x) = \delta^{4}(x)$, the rate in energy given by Eq.~\eqref{eq.E3} becomes 
\begin{equation}
\frac{\D E}{\D z^0}=\frac{\gamma\alpha^2  V}{2}\left(\partial_{\tau} \mathcal N(\tau,0)\mathcal D(0,0)+\mathcal N(0,0)\partial_{\tau}\mathcal D(\tau,0)\right)\bigr|_{\tau=0}.
\end{equation}
From the standard equal-time commutation relations of $\hat\varphi$, which enter the definitions of $\mathcal{N}$ and $\mathcal{D}$, we get $\mathcal{D}(0,0)=0$ and {$\partial_{\tau}\left.\mathcal{D}(\tau,0)\right|_{\tau=0} = -\partial_{z^0}\left.\mathcal{D}(x,z)\right|_{x=z}=\delta(\mathbf{0})$. One can also obtain this result directly from the expression of $\mathcal{D}(x)$ in Eq.~\eqref{eq.defND2}}. This implies that the increase in energy diverges as 
\begin{equation}\label{eq.E3.Markovian}
\frac{\D E}{\D z^0}=\delta(\mathbf{0})\frac{\gamma\alpha^2  V \Tr{\hat \varphi^2(0)\hat \rho(0)}}{2}.
\end{equation}
Here we have used the relation $\mathcal{N}(0,0) =  \Tr{\hat \varphi^2(0)\hat \rho(0)}$. Note that, in Eq.~\eqref{eq.E3.Markovian}, in addition to the standard QFT divergence (of the type one might encounter in computing $\bra{\text{in}}\hat{\varphi}^2\ket{\text{in}}$), there is a multiplicative diverging factor $\delta(\mathbf{0})$. Even if the standard QFT divergence is removed by standard procedures, for example by imposing a normal ordering on the energy rate ($:\D/\D z^0  \int\D\z\,\hat{\mathcal H}^{(0)}(z)=0:$), Eq.~\eqref{eq.E3.Markovian} still remains divergent due to the multiplicative factor $\delta(\mathbf{0})$. This difficulty with relativistic Markovian models, which has already been pointed out in previous works \cite{Pearle1990,Pearle2007,Pearle2009,jones2021mass, zadra2023translation,petrosyan2022relativistic}, serves as a motivation to work with a more generic, non-Markovian noise function.

\section{MCC with normal ordering} \label{App:AppendixC}
In the main text MCC was shown to be satisfied to all orders in $\xi$, by showing that  
\begin{align}\label{seq.Micro}
\fint \xi_n...\xi_1\comm{{\comm{\hat Q_n,...}{\comm{\hat Q_1}{\hat {O}^{(0)}(\tilde{z}_2)}}}}{\hat {O}^{(0)}(\tilde{z}_1)}=0,
\end{align}
whenever $|\tilde{z}_2-\tilde{z}_1|<0$, and hence $|z_2-z_1|<0$. Here, the shorthands $\xi_n$ and $\hat{Q}_1$ stand for $\xi(x_n)$ and $\hat{Q}(x_1)$ respectively, i.e., the quantities evaluated at their respective integration variables.
Here we will show that this condition remains unchanged if the time-evolved operator $\hat{O}(z)$ is normal ordered. That is 
\begin{align}\label{seq.MicroNO}
\fint \xi_n\dots\xi_1\comm{:\!{\comm{\hat Q_n,\dots}{\comm{\hat Q_1}{\hat {O}^{(0)}(\tilde{z}_2)}}}\!:}{\hat {O}^{(0)}(\tilde{z}_1)} =0, 
\end{align}
for $|\tilde{z}_2-\tilde{z}_1|<0$.
This is straightforward to see for operators $\hat{O}^{(0)}(\tilde{z}_2)$ that are linear or at most quadratic in  creation and annihilation operators, such as $\hat{O}^{(0)}\propto \hat{\varphi}$ or  $\hat{O}^{(0)}\propto \hat{\varphi}^2$, or {the free Hamiltonian $\hat{H}_0$}. This is because the collapse operator $\hat Q(x)=\frac{1}{2}\alpha\hat \varphi^2(x)$ is quadratic, and therefore the commutator $\comm{\hat Q(x_1)}{\hat {O}^{(0)}(\tilde{z}_2)}$ would be of the same order in creation and annihilation operators as $\hat{O}^{(0)}$.  Thus, the additional normal ordering can only add a constant, leaving the outermost commutator with $\hat{O}^{(0)}(\tilde{z}_1)$ in Eq.~\eqref{seq.MicroNO} unchanged, when $\hat{O}^{(0)}$ is at most quadratic in creation and annihilation operators.

For higher order operators $\hat{O}^{(0)}$, such as $\hat{O}^{(0)}(x)\propto \hat{\varphi}^4(x)$, the value of the commutator would in general be different with normal ordering when $|\tilde{z}_2-\tilde{z}_1|\geq0$. However, it can be seen with the help of Wick's theorem that the commutator would still be zero when $|z_2-z_1|<0$. 
To show this, we first point out that for ${\comm{\hat Q(x_n),...}{\comm{\hat Q(x_1)}{\hat {O}^{(0)}(\tilde{z}_2)}}}$ to be non-zero, $x_n$, $x_{n-1}$,..., $x_1$ must all be in the past lightcone of $\tilde{z}_2$, as argued in the main text. This does not change with the normal ordering that we impose. We now look at one of the many terms, such as
\begin{align}\label{seq.MicroNOSpecific}
\fint \xi(x_n)...\xi(x_1)\comm{:{{\hat Q(x_n)...}{{\hat Q(x_1)}{\hat {O}^{(0)}(\tilde{z}_2)}}}:}{\hat {O}^{(0)}(\tilde{z}_1)}, 
\end{align}
that contributes to the commutator in  Eq.~\eqref{seq.MicroNO}.
The product $:{{\hat Q(x_n)...}{{\hat Q(x_1)}{\hat {O}^{(0)}(\tilde{z}_2)}}}:$ can be written using Wick's theorem as \cite{Greiner1996Ch8}
\begin{align}\label{seq.Wick}
:{{\hat Q(x_n)...}{{\hat Q(x_1)}{\hat {O}^{(0)}(\tilde{z}_2)}}}: =& \mathcal{T}\{{{\hat Q(x_n)...}{{\hat Q(x_1)}{\hat {O}^{(0)}(\tilde{z}_2)}}}\}\nonumber\\
&- \text{all possible contractions}.
\end{align}
The time ordering would simply change the ordering of the operators as $\mathcal{T}\{\hat Q(x_n)...\hat {O}^{(0)}(\tilde{z}_2)\}\rightarrow \hat {O}^{(0)}(\tilde{z}_2)...\hat Q(x_n)$. It does not change the statement that for Eq.~\eqref{seq.MicroNOSpecific} to be non-zero, it is necessary for $\tilde{z}_1$ to be inside the lightcone of at least one of the spacetime points $x_1, x_2,...,x_n,\tilde{z}_2$. Again, since $x_n$, $x_{n-1}$,..., $x_1$ all lie inside the past lightcone of $\tilde{z}_2$, it implies that $\tilde{z}_1$ must lie  inside the past lightcone of $\tilde{z}_2$, for Eq.~\eqref{seq.MicroNOSpecific} to be non-zero. The same argument holds for the terms involving the contractions in Eq.~\eqref{seq.Wick}. Thus,  Eq.~\eqref{seq.MicroNO}, and hence MCC, is satisfied even in the presence of the additional normal ordering that we impose to make the energy rate non-divergent.

\end{document}